\documentclass[prb,aps,twocolumn,superscriptaddress,longbibliography,floatfix]{revtex4-2}
\usepackage[utf8]{inputenc}
\usepackage{graphicx,color}
\usepackage{amsthm}
\usepackage{amsfonts}
\usepackage{algorithmic}
\usepackage{enumerate}
\usepackage{latexsym}
\usepackage{amsmath}
\usepackage{amssymb}
\usepackage{graphicx}
\usepackage[colorlinks=true,citecolor=blue,linkcolor=blue, urlcolor=blue]{hyperref}

\emergencystretch\maxdimen
\begin{document}
	
	\title{Comparison of superconducting pairing in doped cuprates and nickelates within the Hubbard model including the third-nearest neighbor hopping terms}
	
	\author{Yicheng Xiong}
	\affiliation{School of Physics and Astronomy, Beijing Normal University, Beijing 100875, China\\}
	
	\author{Hang Ma}
	\affiliation{School of Physics and Astronomy, Beijing Normal University, Beijing 100875, China\\}
	
	\author{Hongxing Liu}
	\affiliation{School of Physics and Astronomy, Beijing Normal University, Beijing 100875, China\\}
	
	\author{Runyu Ma}
	\affiliation{School of Physics and Astronomy, Beijing Normal University, Beijing 100875, China\\}

	\author{Tianxing Ma}
	\email{txma@bnu.edu.cn}
	\affiliation{School of Physics and Astronomy, Beijing Normal University, Beijing 100875, China\\}
	\affiliation{Key Laboratory of Multiscale Spin Physics (Ministry of Education), Beijing Normal University, Beijing 100875, China\\}
	
	\begin{abstract}
		Within a Hubbard model, we investigate the superconducting pairing behavior of infinite-layer nickelate $\mathrm{NdNiO_2}$ and cuprate superconductors by using the determinant quantum Monte Carlo method. Our focus is on comparing their dominant pairing symmetries.
		The results indicate that the $d_{x^2-y^2}$ pairing interaction is significantly enhanced at low temperatures in both doped nickelates and cuprates, whereas other typical pairing symmetries are effectively suppressed, highlighting the dominance of the $d_{x^2-y^2}$ pairing form. Additionally, we find that the effective pairing interaction for $d_{x^2-y^2}$ pairing in doped nickelates is slightly lower than that in doped cuprates, which may be attributed to the different degrees of Fermi surface warping caused by the third-nearest hopping $t''$. Further studies show that the hole doping and interaction strength have significant effects on the $d_{x^2-y^2}$ pairing interaction within the selected parameter range. The $d_{x^2-y^2}$ pairing interaction is notably weakened when the hole doping increases, whereas it is significantly enhanced with increasing Coulomb interaction strength $U$. This comparative analysis reveals the similarities and differences in the pairing behaviors of doped nickelates and cuprates, which may provide further insights into understanding the superconducting properties of these two classes of materials.
	\end{abstract}
	\maketitle
	
	\section{Introduction}
	Since the discovery of high-$T_c$ superconductivity in doped cuprates \cite{bednorzPossibleHighTcSuperconductivity1986a}, unraveling the mechanism of superconductivity \cite{andersonPhysicsHightemperatureSuperconducting2004,azumaSuperconductivity110Infinitelayer1992,damascelliAngleresolvedPhotoemissionStudies2003,leeDopingMottInsulator2006,scalapinoCommonThreadPairing2012} and the dependence of the transition temperature on material properties have been central issues in condensed matter physics. Comparative studies of materials with similar features but different chemical compositions should be an effective way to understand this problem. Recent breakthroughs in nickelates, such as the achievement of superconducting transition temperatures up to 15 K in Nd$_{1-x}$Sr$_{x}$NiO$_{2}$ \cite{liSuperconductivityInfinitelayerNickelate2019d,sawatzkySuperconductivitySeenNonmagnetic2019}and 9 K in La$_{1-x}$Sr$_{x}$NiO$_{2}$ \cite{osadaNickelateSuperconductivityRareearth2021}, have provided new perspectives for understanding the pairing mechanism in high-temperature superconductors. The 3$d$$^{9}$ electronic configuration of Ni$^{+}$ in NdNiO$_{2}$ resembles that of Cu$^{2+}$ in cuprates \cite{anisimovElectronicStructurePossible1999,leeInfinitelayerMathrmMathrm2004}. The Ni$^{+}$ oxidation state is stabilized in the infinite-layer square-planar $R$NiO$_{2}$ ($R$ = La, Nd, Pr) \cite{haywardSodiumHydridePowerful1999,osadaPhaseDiagramInfinite2020c}, which possesses the same P4/mmm crystal structure as the parent compounds of high-T$_{c}$ cuprates \cite{botanaSimilaritiesDifferencesLaNiO22020}. Superconductivity is also observed when neodymium is substituted with other lanthanides \cite{osadaPhaseDiagramInfinite2020c,osadaNickelateSuperconductivityRareearth2021,zengSuperconductivityInfinitelayerNickelate2022c} and in more complex multilayer structures \cite{panSuperconductivityQuintuplelayerSquareplanar2022b}. These findings reveal an intrinsic similarity between doped nickelates and cuprates.

	Currently, theoretical studies on some nickel-based superconductors have constructed multiorbital models incorporating Ni-3$d$ and R-5$d$ (where R represents rare-earth elements) orbitals to investigate orbital hybridization \cite{PhysRevX.11.011050,heptingElectronicStructureParent2020e,PhysRevB.106.224517} and spin interactions \cite{yang2022self,PhysRevB.101.060504} on the superconducting properties of infinite-layer nickelates. These studies suggest that multiband effects may significantly influence the superconducting pairing behavior of infinite-layer nickelates. However, other researchers have proposed a single-band Hubbard model to describe and capture the fundamental characteristics of the normal and superconducting states of infinite-layer nickelates \cite{PhysRevLett.125.147003,PhysRevLett.130.166002,WOS:001032441500016}. In doped nickelates and cuprates, the low-energy degrees of freedom are predominantly determined by the 3$d$ orbitals. However, their electronic structures exhibit some differences. First, unlike doped cuprates, the energy difference between O-2$p$ and Ni-3$d$ in doped nickelates is much greater than the Coulomb interaction between Ni-3$d$ electrons \cite{sawatzkySuperconductivitySeenNonmagnetic2019,zhangTypeIITextEnsuremath2020,jiangCriticalNatureNi2020a,huTwobandModelMagnetism2019a}, leading to the direct entry of additional holes into Ni-3$d$ orbitals rather than the formation of Zhang-Rice singlets \cite{zhangEffectiveHamiltonianNickelate2020}. Therefore the influence of the O-2$p$ degree of freedom can be safely neglected. Second, although electron pockets A and $\Gamma$ formed by Nd-5$d_{xy}$ and Nd-5$d_{z^2}$ orbitals exist, these electron pockets hardly hybridize with the Ni-3$d_{x^2-y^2}$ band  \cite{kitatani2020nickelate,di2024unconventional} in the low doping region (Sr doping concentration $<24\%$ \cite{PhysRevLett.125.147003}). The A pocket can be considered a pure hole reservoir, accommodating some of the holes induced by Sr doping. For conductivity and other transport properties, the A pocket may also be relevant, but superconductivity should originate from the correlated 3$d_{x^2-y^2}$ band that has almost no coupling with the A pocket \cite{held2022phase}. Using the single-band Hubbard model derived from these observations, the key features of the superconducting dome in Sr-doped NdNiO$_2$ were accurately predicted before recent experimental realization from defect-free films  \cite{PhysRevLett.125.147003,WOS:001032441500016}. An experimental study \cite{PhysRevLett.133.211003} also revealed the crucial role of the hydride reduction process in regulating Ni-3$d$ orbital polarization. X-ray absorption spectroscopy analysis revealed significant linear dichroism in the superconducting samples, which was attributed mainly to the Ni-3$d_{x^2-y^2}$ orbitals, indicating that infinite-layer nickelates exhibit single-band properties. These research findings support the use of the single-band model as a reasonable starting point for studying infinite-layer nickelates.
	
	Many theoretical studies have revealed that $d$-wave pairing is dominant in nickelates \cite{zhangTypeIITextEnsuremath2020,zhang2021theory}. Furthermore, various quantum Monte Carlo (QMC) methods have been widely used to study $d$-wave superconductivity, including exploring the competition between nodal $d$-wave superconductivity and antiferromagnetic order \cite{PhysRevLett.126.217002}, investigating the suppression of $d$-wave pairing in cuprates and nickelates due to rotational symmetry breaking \cite{PhysRevB.110.L201116}, and revealing the influence of the period of charge stripes on $d$-wave pairing \cite{chen2024charge}. Understanding the factors that influence the superconducting pairing behavior is also a major research focus in the field of superconductivity. Currently, theoretical studies that introduce the next-nearest-neighbor hopping \( t' \) into the Hubbard model---using methods such as exact diagonalization, density matrix renormalization group (DMRG), and QMC---have explored the impact of \( t' \) on electronic properties, such as impact ionization effects, charge stripes, and \( d \)-wave pairing symmetry \cite{PhysRevB.102.245125,xu2024coexistence,jiang2019superconductivity,yang2020quantum}. These studies suggest that \( t' \) can significantly affect the superconducting pairing symmetry in the Hubbard model. Motivated by this, using the determinant quantum Monte Carlo (DQMC) method, we further investigated the influence of the third-nearest hopping \( t'' \) on the dominant pairing symmetry when comparing superconducting pairing properties in cuprates and nickelates under varying doping levels and interaction strengths. We find that, with increasing Coulomb interaction strength \( U \), the pairing susceptibility of the \( d_{x^2 - y^2} \) channel is significantly increased, indicating that electronic correlations play a crucial role in promoting \( d_{x^2 - y^2} \) pairing. Furthermore, we find that the third-nearest hopping \( t'' \) has a suppressive effect on the \( d_{x^2 - y^2} \) pairing channel.
	
	\section{Model and methods}
	In the single-band Hubbard model, the nickel-square lattice Hamiltonian can be written as
	\begin{align}
		H &= H_{K} + H_{\mu} + H_{V}, \label{eq:H_total}\\
		H_{K} &= -t\sum_{\langle\mathbf{i},\mathbf{j}\rangle,\sigma}(c_{\mathbf{i}\sigma}^{\dagger}c_{\mathbf{j}\sigma}+c_{\mathbf{j}\sigma}^{\dagger}c_{\mathbf{i}\sigma}) \nonumber\\
		&\quad -t'\sum_{\langle\mathbf{i},\mathbf{j}\rangle,\sigma}(c_{\mathbf{i}\sigma}^{\dagger}c_{\mathbf{j}\sigma}+c_{\mathbf{j}\sigma}^{\dagger}c_{\mathbf{i}\sigma}) \nonumber\\
		&\quad -t''\sum_{\langle\mathbf{i},\mathbf{j}\rangle,\sigma}(c_{\mathbf{i}\sigma}^{\dagger}c_{\mathbf{j}\sigma}+c_{\mathbf{j}\sigma}^{\dagger}c_{\mathbf{i}\sigma}), \label{eq:H_K}\\
		H_{\mu} &= -\mu\sum_{\mathbf{i}}(n_{\mathbf{i}\uparrow}+n_{\mathbf{i}\downarrow}), \label{eq:H_mu}\\
		H_{V} &= U\sum_{\mathbf{i}}(n_{\mathbf{i}\uparrow}-\frac12)(n_{\mathbf{i}\downarrow}-\frac12). \label{eq:H_V}
	\end{align}
	For the copper-oxygen plane, we take the nearest hopping $t$ as the unit of energy (set to 1), the second-nearest hopping $t' = -0.25t$, and the third-nearest hopping $t'' = 0$. For the nickel-oxygen plane, we take $t = 1$, $t' = -0.25t$, and $t'' = 0.12t$. These parameters are derived from density functional theory (DFT) \cite{PhysRevLett.130.166002,marzariMaximallyLocalizedWannier2012,blahaWIEN2kAPWProgram2020,perdewGeneralizedGradientApproximation1996}.
	The fermionic operators $c_{i\sigma}^\dagger$ and $c_{i\sigma}$ represent the creation and annihilation of an electron with spin $\sigma$($\sigma={\uparrow, \downarrow}$) at site $i$, respectively, satisfying the anticommutation relations. The particle number operator $n_{i\sigma}=c_{i\sigma}^\dagger c_{i\sigma}$ gives the number of electrons with spin $\sigma$ at site $i$. The chemical potential $\mu$ is used to tune the electron density of the system. By changing the value of $\mu$, one can control the doping level of the system.
	\begin{figure}[tbp]
		\includegraphics[scale=0.19]{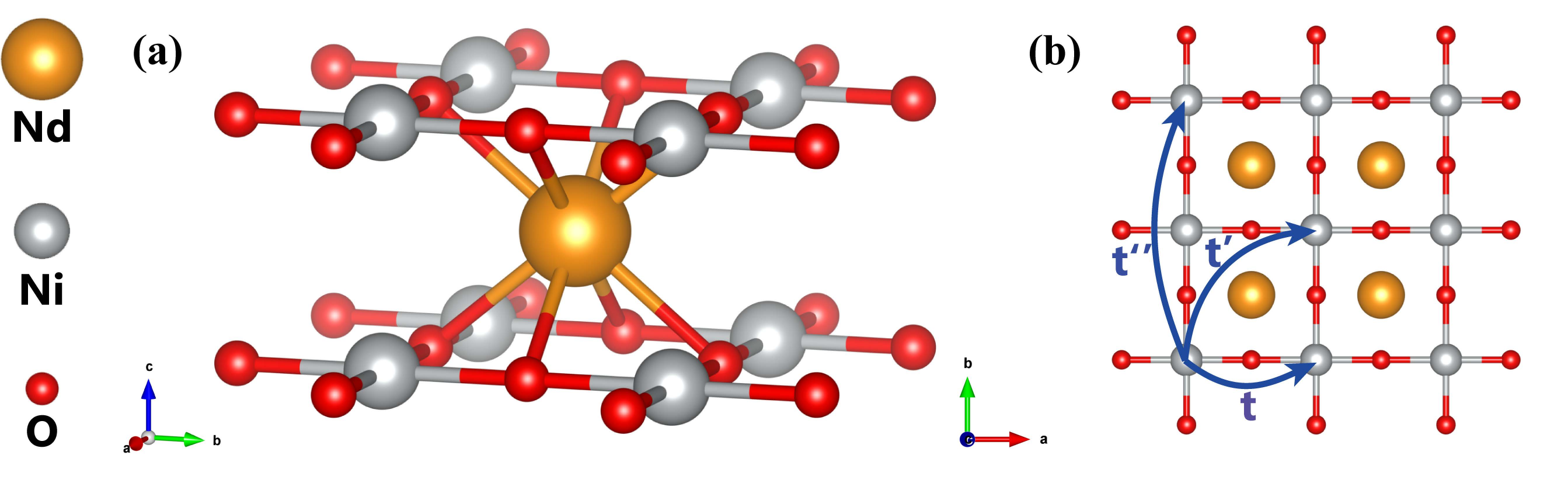}
		\caption{(a) Crystal structure of NdNiO$_{2}$ compound, which consists of Nd, Ni, and O atoms, exhibiting a distinct layered arrangement. (b) Top view of the lattice structure of NdNiO$_{2}$ compound, where the three blue arrows represent the nearest, second-nearest, and third-nearest hopping modes of electrons in the lattice.}
		\label{1}
	\end{figure}

We employed the DQMC method to simulate the lattice illustrated in Fig. \ref{1}(b) under periodic boundary conditions. By utilizing the Suzuki-Trotter decomposition, we express the partition function $Z$ via the Suzuki-Trotter decomposition, which introduces the imaginary-time interval $\Delta \tau$. A discrete Hubbard-Stratonovich transformation is then applied to decouple the interaction term, facilitating the efficient computation of the fermionic trace. This transformation eliminates the fermionic degrees of freedom by introducing auxiliary Ising fields, which are initially generated randomly and depend on both spatial sites and imaginary time. These fields are subsequently updated through local spin flips, with acceptance probabilities determined by the Metropolis algorithm. A single DQMC sweep consists of traversing the entire space-time lattice to update the auxiliary field variables.

To equilibrate the system, we performed 8,000 warm-up sweeps, after which the physical observables stabilized. We then conducted 30,000 sweeps for measurements. For certain data points, especially those affected by the sign problem (see Appendix Fig. \ref{FIG9}), we increased the number of measurement sweeps to 120,000, dividing them into 20 bins for statistical averaging. Throughout this process, we set \(\Delta \tau = 0.125\) to ensure that the Suzuki-Trotter error was smaller than the statistical sampling error \cite{PhysRevB.85.125127}. For the technical details of the DQMC method, please refer to Refs.~\cite{hirschDiscreteHubbardStratonovichTransformation1983,whiteAttractiveRepulsivePairing1989a,blankenbeclerMonteCarloCalculations1981a,maQuantumMonteCarlo2013c,maLocalizationInteractingDirac2018b}
	
\begin{figure}[tbp]
		\includegraphics[scale=0.75]{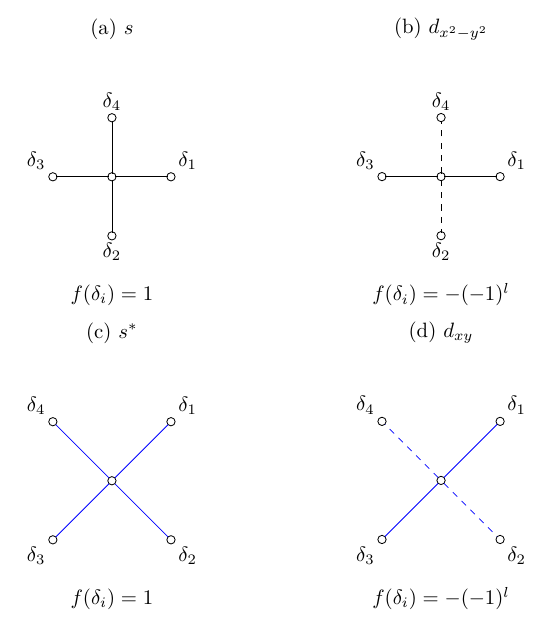}
		\caption{Five pairing forms possess different symmetries and phases. Specifically, (a) is the $s-$wave, (b) is the $d_{x^2-y^2}-$wave, (c) is the extended $s-$wave, and (d) is the $d_{xy}-$wave.}
		\label{2}
	\end{figure}

	To investigate the superconductivity in doped nickelates and cuprates, we define the superconducting pairing susceptibility as
	\begin{equation}P_\alpha=\frac1{N_s}\sum_{\mathrm{i,j}}\int_0^\beta d\tau\langle\Delta_\alpha^\dagger(\mathbf{i},\tau)\Delta_\alpha(\mathbf{j},0)\rangle,\label{pairs}\end{equation}
	where the superconducting order parameter $\Delta_\alpha^\dagger(\mathbf{i})$ is given by
	\begin{equation}\Delta_\alpha^\dagger(\mathbf{i})=\sum_lf_\alpha^\dagger(\delta_l)(c_{\mathbf{i}\uparrow}c_{\mathbf{i}+\delta_l\downarrow}-c_{\mathbf{i}\downarrow}c_{\mathbf{i}+\delta_l\uparrow})^\dagger \label{delta}\end{equation}
	The phase factors $\delta_{l}$ associated with different pairing forms determine superconducting pairing symmetry $\alpha$. In the square lattice, we investigate the pairing forms shown in Figs. \ref{2}(a)-\ref{2}(d). The structure factors of these pairing symmetries can be expressed as follows:
	\begin{equation}\begin{aligned}
			s\mathrm{-wave:}f_{s}(\delta_{l})& =1(\delta_l=(\pm(x,\pm\hat{y})),  \\
			d_{x^{2}-y^{2}}\mathrm{-wave}:f_{d_{x^{2}-y^{2}}}(\delta_{l})& =1(\delta_l=(\pm\hat{x},0)),  \\
			\mathrm{and~}f_{d_{x^{2}-y^{2}}}(\delta_{l})& =-1(\delta_{l}=(0,\pm\hat{y})),  \\
			\mathrm{extended}\, s-\mathrm{wave}:f_{s^{*}}(\delta_{l})& =1(\delta_{l}=\pm(x,\pm\hat{y})),  \\
			d_{xy}\mathrm{-wave}:f_{d_{xy}}(\delta_{l})& =1(\delta_l=\pm(-\hat{x},\hat{y})),  \\
			\mathrm{and} f_{d_{xy}}(\delta_{l})& =-1(\delta_l=\pm(\hat{x},\hat{y})).  \\
	\end{aligned}\end{equation}
	To determine the dominant pairing symmetry in doped nickelates and cuprates, we introduce the effective pairing interaction $\bar{P}{\alpha}=P{\alpha}-\tilde{P}{\alpha}$ \cite{maQuantumMonteCarlo2013c,huangAntiferromagneticallyOrderedMott2019g,chenAntiferromagneticFluctuationsDominant2022d,chenStripeOrderManipulated2024b}, where the uncorrelated single-particle contribution $\tilde{P}{\alpha}$ can be obtained by replacing $\langle c_{i\downarrow}^{\dagger}c_{j\downarrow}c_{i+\delta_l\uparrow}^{\dagger}c_{j+\delta_{l'}\uparrow}\rangle$ with $\langle c_{i\downarrow}^{\dagger}c_{j\downarrow}\rangle\langle c_{i+\delta_l\uparrow}^{\dagger}c_{j+\delta_{l'}\uparrow}\rangle$ in Eq. (\ref{delta}). A positive effective pairing interaction indicates that the superconducting pairing form is driven by electron-electron correlations, whereas a negative value suggests that this superconducting pairing channel may be suppressed by other pairing channels.
	
	\begin{figure}[tbp]
		\includegraphics[scale=0.425]{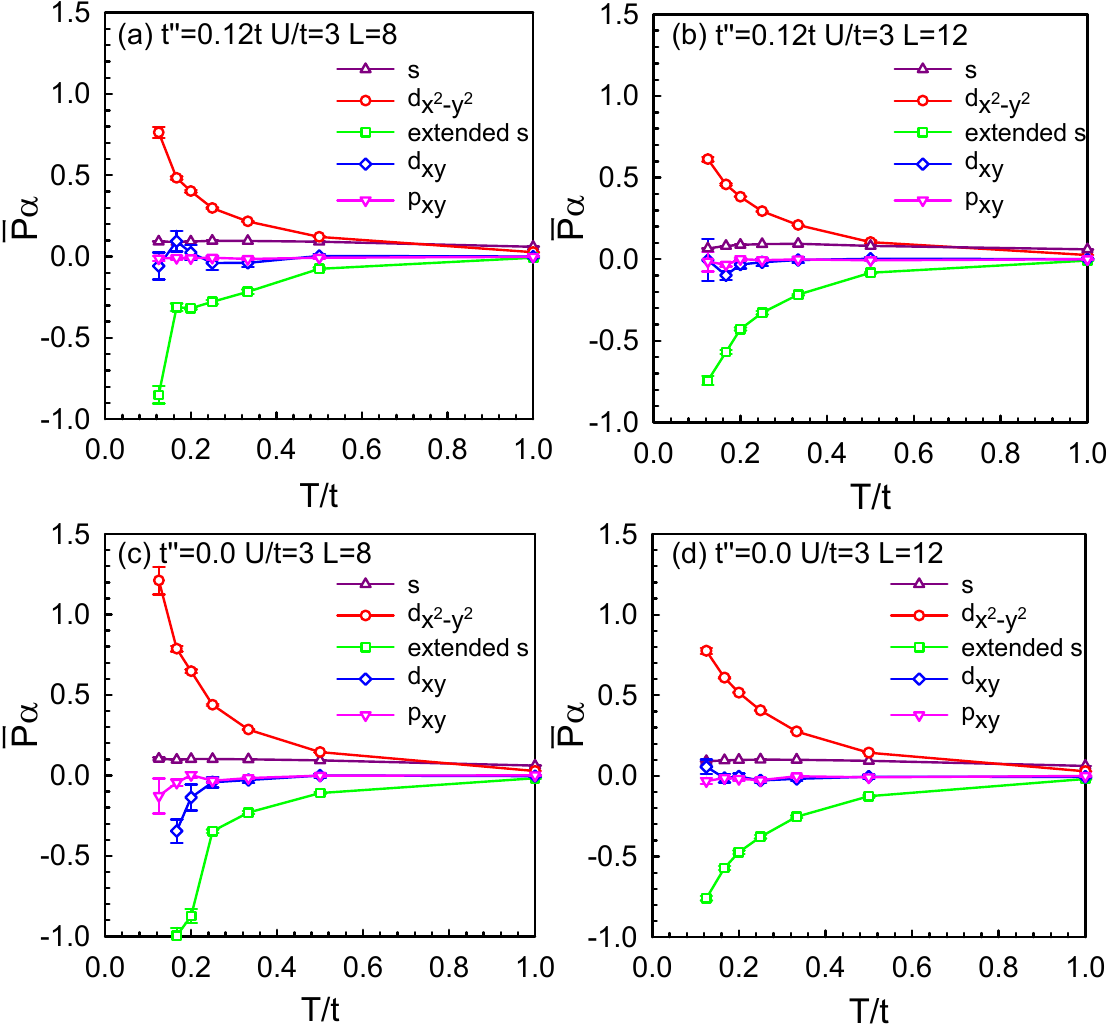}
		\caption{[(a) and (b)] Temperature dependence of the effective pairing interaction for different candidate pairing symmetries in doped nickelate with lattice sizes of $8\times8$ and $12\times12$, respectively, at $U/t=3$ and electron density $\langle n\rangle=0.8$. [(c) and (d)] Corresponding results for doped cuprates under the same conditions.}
		\label{3}
	\end{figure}
	
	\section{Results}
	Our main simulations are performed on a two-dimensional square lattice with $8 \times 8$ sites, and further results are also checked on a lattice with  $12 \times 12$ sites, which is fairly large for the DQMC method. We investigated the temperature dependence of the effective pairing interactions for various candidate pairing symmetries with these lattices. As shown in Fig. \ref{3}, at low temperatures, we observed a significant increase in the effective pairing interaction of the $d_{x^2-y^2}$ symmetry in both doped nickelates and cuprates. This suggests that the $d_{x^2-y^2}$ pairing symmetry may play a crucial role in these strongly correlated electron systems. However, notably, the $d_{x^2-y^2}$ effective pairing interaction in doped nickelates is slightly lower in magnitude than that in doped cuprates; this discrepancy may be attributed to different degrees of Fermi surface warping caused by distinct tight-binding parameters. Moreover, we examined other typical superconducting pairing symmetries, namely, the $s-$wave, extended $s-$wave, $d_{xy}-$wave, and $p_{xy}-$wave. The results indicate that the effective pairing interactions for these channels are either close to zero or negative; this implies that the $d_{x^2-y^2}$ superconducting pairing symmetry has an intrinsic competitive advantage over other pairing channels, effectively suppressing the formation of other superconducting states. This finding provides strong evidence for the dominance of $d_{x^2-y^2}$ pairing symmetry near optimal doping.

	\begin{figure}[tbp]
		\includegraphics[scale=0.425]{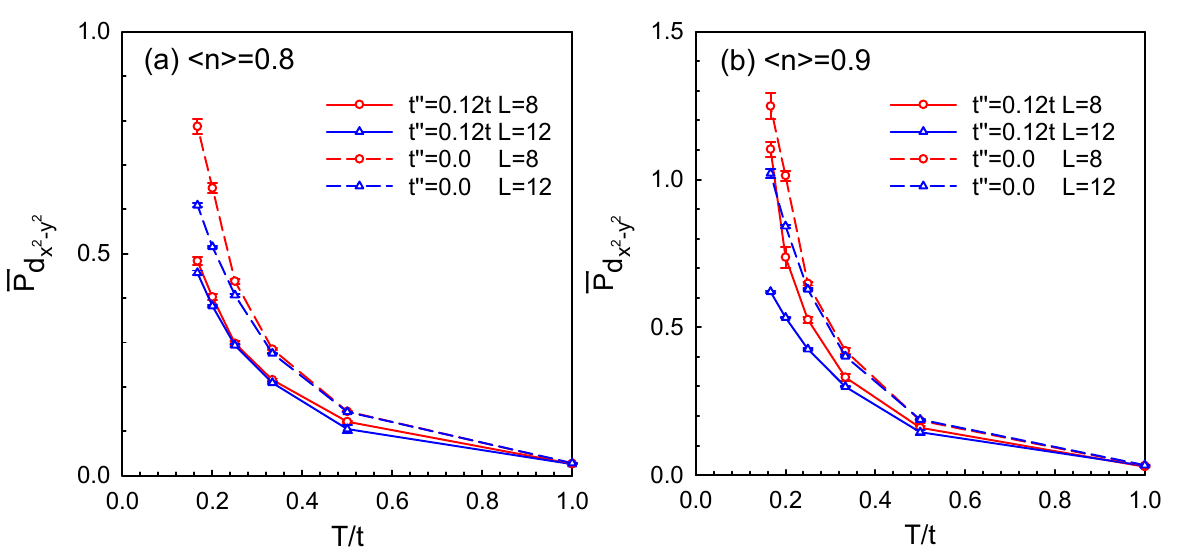}
		\caption{Temperature dependence of the effective $d_{x^2-y^2}$ pairing interaction for doped nickelates (solid lines) and cuprates (dashed lines) at different lattice sizes, where circles represent $L=8$ and upward triangles represent $L=12$. The calculations are performed with $U/t=3$ and the electron density $\langle n \rangle=0.8,0.9$.}
		\label{4}
	\end{figure}
	
	We also investigated the influence of the lattice size on the computational results. As shown in Fig. \ref{4}, the $d_{x^2-y^2}$ pairing channel has a weak dependence on the lattice size of both  doped cuprates and nickelates. The effective pairing interaction remains nearly constant with varying lattice size, indicating the robustness of the $d_{x^2-y^2}$ wave dominance.
	
\begin{figure}[tbp]
		\includegraphics[scale=0.425]{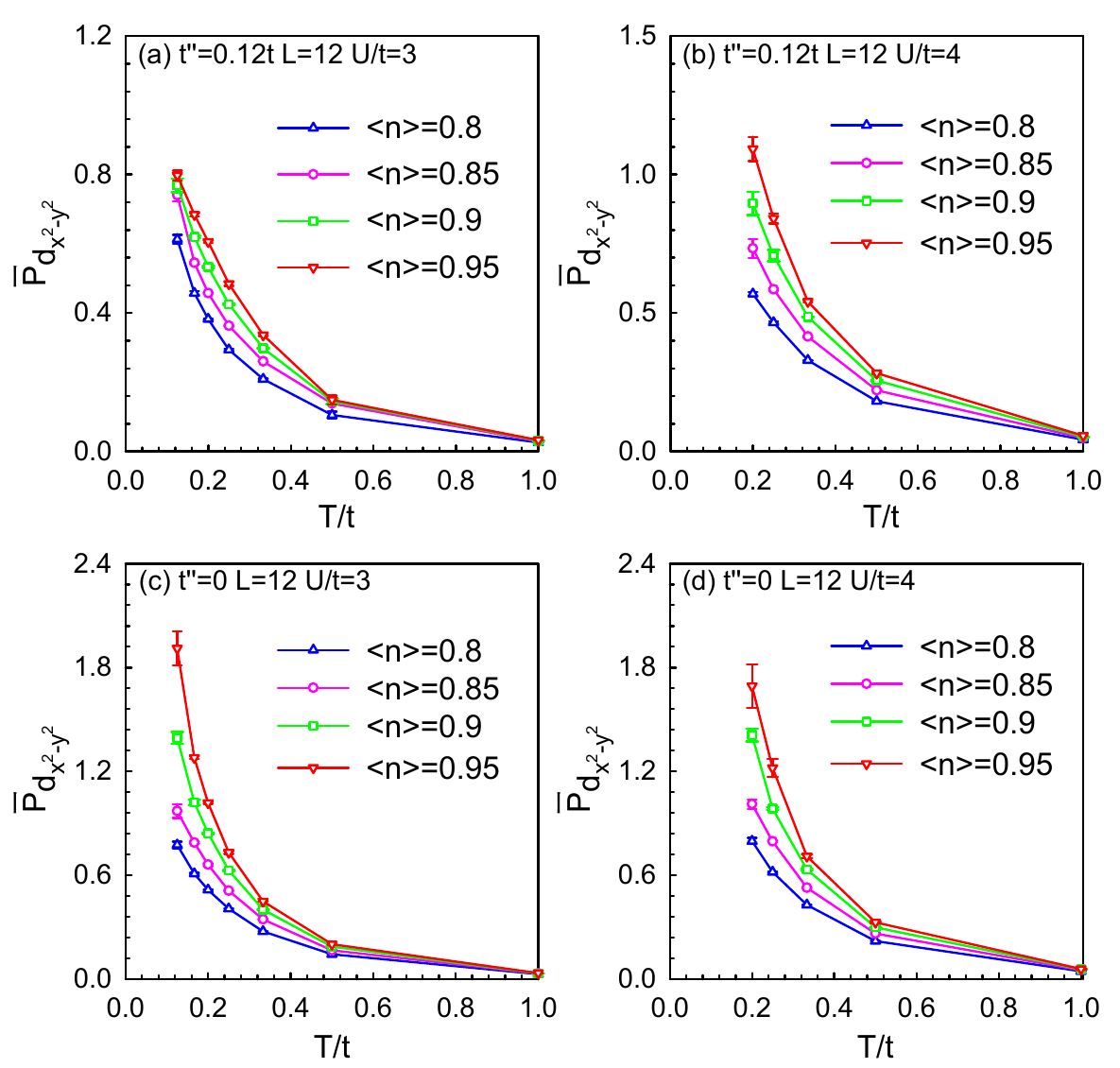}
		\caption{Temperature dependence of the effective $d_{x^2-y^2}$ pairing interaction for doped nickelates and cuprates with different hole doping amounts. [(a) and (b)] Results for doped nickelates with a lattice size of $L=12$ and Coulomb interaction strengths of $U/t=3$ and 4, respectively. [(c) and (d)] Corresponding results for doped cuprates under the same conditions.}
		\label{5}
	\end{figure}

	To systematically investigate the influence of hole doping on the temperature dependence of the $d_{x^2-y^2}$ effective pairing interaction, in Fig. \ref{5}, we present the evolution of the $d_{x^2-y^2}$ effective pairing interaction as a function of temperature for doped cuprate and nickelate compounds at $U/t=3, 4$ and electron filling $\langle n\rangle=0.8, 0.85$, 0.9, 0.95. Within the studied parameter range, as the electron filling gradually deviates from half-filling, the $d_{x^2-y^2}$ effective pairing interaction exhibits a significant decreasing trend in both materials, which is especially more pronounced in the low-temperature region. This indicates that the superconducting pairing has a strong dependence on electron filling, which is a common feature of doped cuprates and nickelates. This may originate from the strong Coulomb repulsion between $d$-orbital electrons in the CuO$_2$ and NiO$_2$ planes, leading to the formation of a Mott insulating state near half-filling. Moderate hole or electron doping can destroy the Mott insulating state, introduce charge carriers, and thus realize high-temperature superconductivity. 
	
\begin{figure}[tbp]
		\includegraphics[scale=0.425]{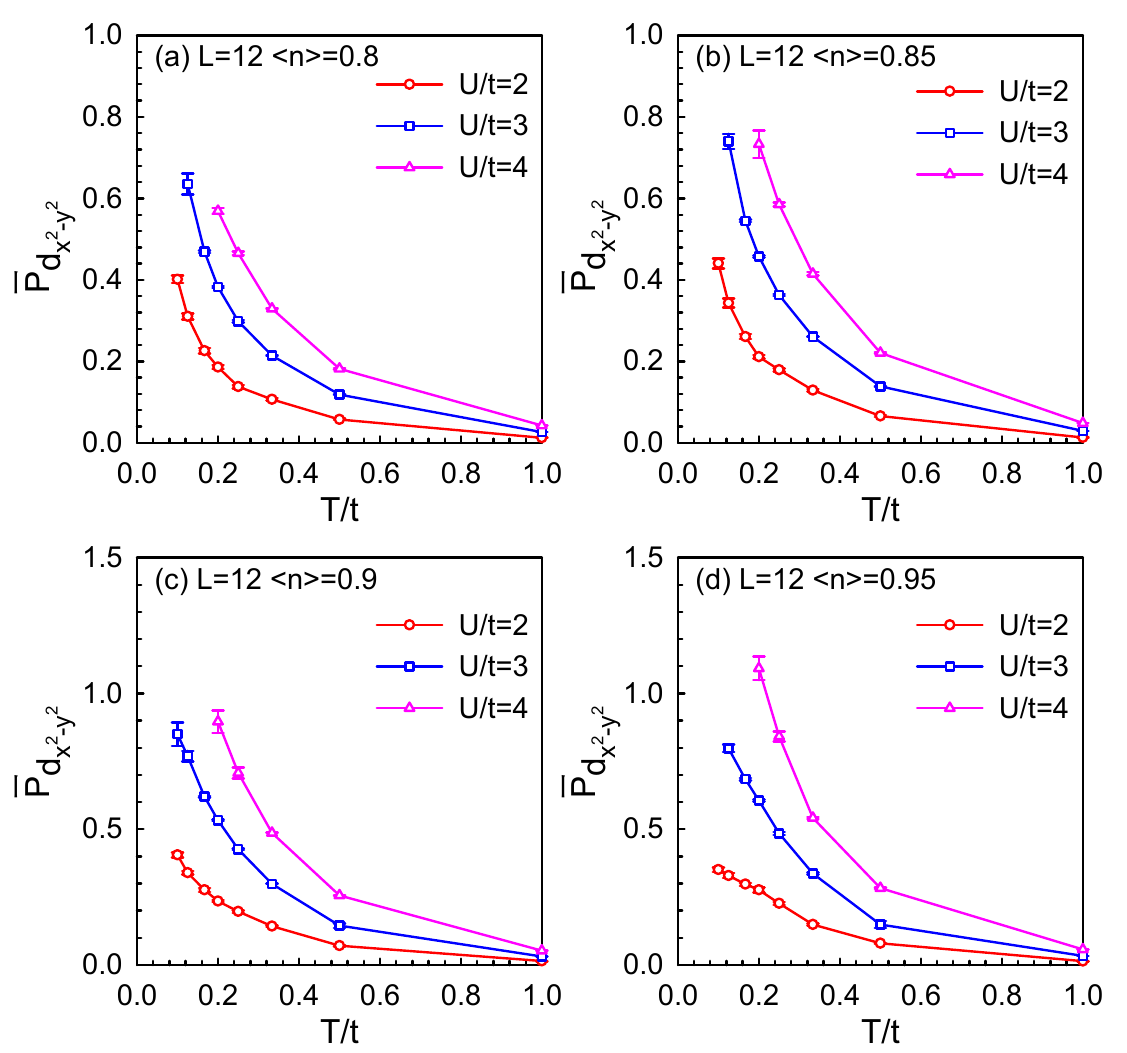}
		\caption{Temperature dependence of the $d_{x^2-y^2}$ effective pairing interaction $\bar{P}_{d_{x^2-y^2}}$ in doped nickelates at different electron densities $\langle n\rangle$: (a) $\langle n\rangle=0.8$, (b) $0.85$, (c) $0.9$, and (d) $0.95$. The curves correspond to various Coulomb interaction strengths $U$. The calculations are performed on a lattice size of $L=12$.}
		\label{6}
	\end{figure}

\begin{figure}[tbp]
		\includegraphics[scale=0.425]{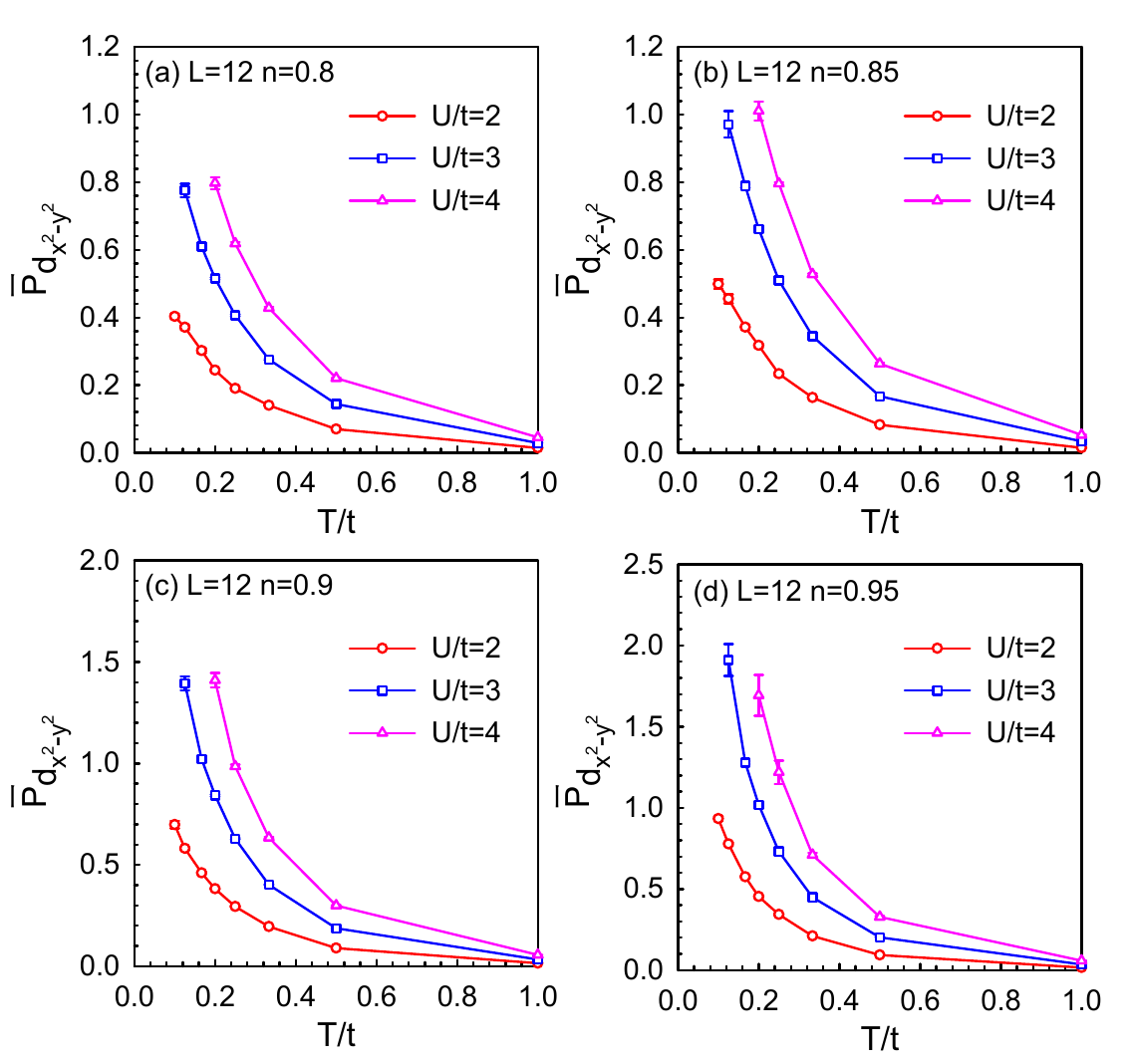}
		\caption{Temperature dependence of the $d_{x^2-y^2}$ effective pairing interaction $\bar{P}_{d_{x^2-y^2}}$ in doped cuprates at different electron densities $\langle n\rangle$: (a) $\langle n\rangle=0.8$, (b) $\langle n\rangle=0.85$, (c) $\langle n\rangle=0.9$, and (d) $\langle n\rangle=0.95$. The curves correspond to various Coulomb interaction strengths $U$. The calculations are performed on a lattice size of $L=12$.}
		\label{7}
	\end{figure}

	Figures \ref{6} and \ref{7} clearly demonstrate the impact of the Coulomb interaction strength $U$ on the effective pairing interaction $\bar{P}_{d_{x^2-y^2}}$ in the $d_{x^2-y^2}$ pairing channel for doped nickelates and cuprates. As $U$ gradually increases, the $\bar{P}_{d_{x^2-y^2}}$ of both materials exhibits a significant increasing trend, indicating that electronic correlation effects play a crucial role in promoting $d_{x^2-y^2}$ superconducting pairing. It is noteworthy that in the low-temperature region, the $\bar{P}_{d_{x^2-y^2}}$ of doped cuprates shows a more pronounced divergence behavior, especially at larger $U$ values. This divergence trend foreshadows the formation of $d_{x^2-y^2}$ superconducting pairing. In contrast, the divergence of $\bar{P}_{d_{x^2-y^2}}$ in doped nickelates at low temperatures is less evident, suggesting that their $d$-wave superconducting pairing may not be as stable and prominent as in doped cuprates.
	
\begin{figure}[tbp]
		\includegraphics[scale=0.425]{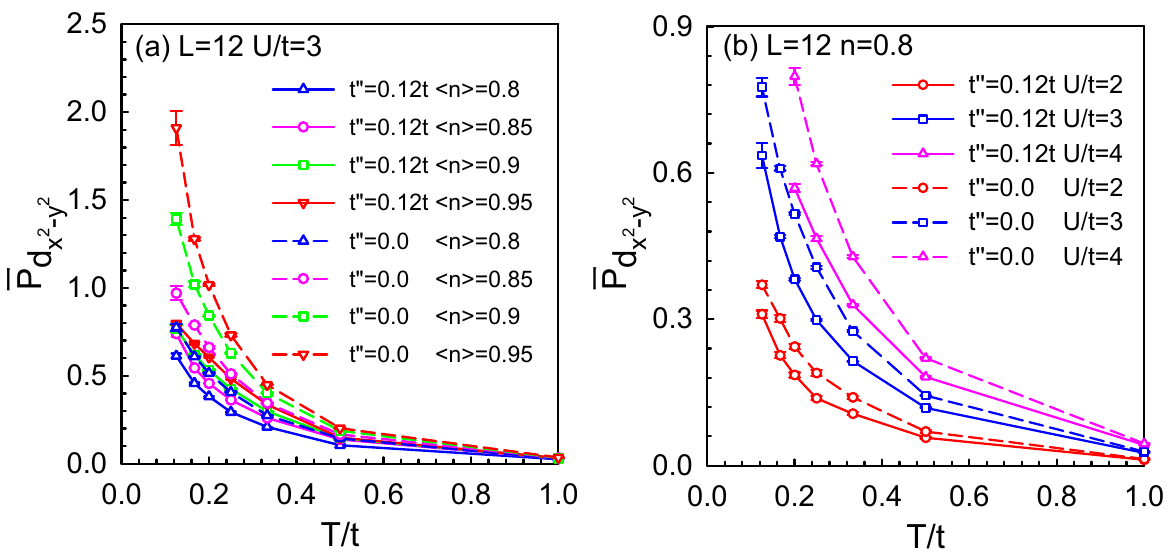}
		\caption{(a) Temperature dependence of the $d_{x^2-y^2}$-wave pairing susceptibility $\bar{P}_{d_{x^2-y^2}}$ for different $t''$ values at lattice size $L=12$, interaction strength $U/t=3$, and electron densities $\langle n\rangle=0.8, 0.85, 0.9, 0.95$. (b) Temperature dependence of the $d_{x^2-y^2}$-wave pairing susceptibility $\bar{P}_{d_{x^2-y^2}}$ for different $t''$ values at lattice size $L=12$, electron density $\langle n\rangle=0.8$, and interaction strengths $U/t=2, 3, 4$.}
		\label{8}
	\end{figure}

	Finally, we investigated the competitive advantage of the $d_{x^2 - y^2}$ wave pairing channel under different $t''$ values. As shown in Fig. \ref{8}(a), when $U/t=3$ and $\langle n \rangle = 0.8, 0.85, 0.9, 0.95$ are fixed, and $t''$ is increased, the effective pairing interaction of the $d_{x^2 - y^2}$ wave exhibits a weakening effect. Similarly, as shown in Fig. \ref{8}(b), when $\langle n \rangle =0.8$ and $U/t =2, 3, 4$ are fixed, and $t''$ is increased, the pairing susceptibility of the $d_{x^2 - y^2}$ wave also shows a weakening effect. This indicates that within the selected parameter range, $t''$ has a certain inhibitory effect on the $d_{x^2 - y^2}$ superconducting pairing channel. This result reveals the important role of the third hopping in regulating the $d_{x^2 - y^2}$ wave superconducting pairing strength for doped nickelates and cuprates. This may be because reducing $t''$ correspondingly reduces the warping of the Fermi surface. Under optimal parameter conditions, as the Fermi surface flattens, superconductivity is enhanced  \cite{PhysRevLett.130.166002}. For a detailed discussion of this view, please see Appendix Fig. \ref{FIG12}.
	
It is worth noting that although \( t'' \) is small in magnitude compared with the nearest-neighbor hopping \( t \), its relative difference between nickelates and cuprates still has a significant effect on superconducting properties as shown in Fig. \ref{FIG11}. Our main point remains that this difference in \( t'' \) is an important factor leading to the differences in superconducting behavior between the two types of materials. Nevertheless, we also recognize that further theoretical and experimental investigations are necessary to clarify its precise relevance in nickelates.
	
	\section{Summary}
	We constructed two-dimensional square lattice models of varying lattice sizes to study the temperature dependence of the effective pairing interactions for different candidate pairing symmetries. The results show that at low temperatures, the $d_{x^2-y^2}$ pairing symmetry dominates in both doped nickelates and cuprates, with a significant increase in the effective pairing interaction, indicating that $d_{x^2-y^2}$ pairing plays a crucial role in these strongly correlated systems. However, the magnitude of the $d_{x^2-y^2}$ effective pairing interaction in doped nickelates is slightly smaller than that in doped cuprates, possibly due to differences in their tight-binding parameters, leading to variations in Fermi surface warping. We also investigated the effect of hole doping on the $d_{x^2-y^2}$ effective pairing interaction. The results indicate that as the filling deviates from half-filling, the $d_{x^2-y^2}$ effective pairing interaction significantly weakens in the low-temperature region, suggesting a dependence of superconducting pairing on electron filling. Additionally, with increasing Coulomb interaction strength $U$, the $d_{x^2-y^2}$ pairing interaction is significantly enhanced, indicating that electron correlation effects play a key role in promoting $d_{x^2-y^2}$ superconducting pairing. Finally, we found that the third hopping $t''$ has a certain inhibitory effect on the $d_{x^2-y^2}$ wave pairing channel.
	
	\noindent
	\underline{\it Acknowledgments}
	This work was supported by the Beijing Natural Science Foundation (No. 1242022) and the Guangxi Key Laboratory of Precision Navigation Technology and Application, Guilin University of Electronic Technology (No. DH202322). The numerical simulations in this work were performed at the HSCC of Beijing Normal University.
	
	\section*{Appendix}
Our selection of thermalization sweeps and measurement sweeps is based on systematic convergence tests and practical considerations of computational efficiency. These choices strike a balance between ensuring accurate and reliable results and maintaining feasible computational demands. To characterize the sign problem, we calculate the average fermionic sign $\langle \operatorname{sign}\rangle$ as the ratio between the integral of the product of the upspin and downspin determinants and the integral of their absolute value product:
	\begin{equation}\langle \operatorname{sign}\rangle = \frac{\sum_{\chi}\det M_{\uparrow}(\chi)\det M_{\downarrow}(\chi)}{|\sum_{\chi}\det M_{\uparrow}(\chi)\det M_{\downarrow}(\chi)|}\end{equation}
	where $M_{\sigma}(\chi)$ represents the matrix for each spin specie. As shown in Fig. ref{FIG9}, $\langle \operatorname{sign} \rangle$ remains significantly large until the temperature $T$ decreases to $0.1667t$. Therefore the conclusion at $U = 3$ and $n = 0.8$ is reliable at high temperatures. When $\langle \operatorname{sign} \rangle$ is relatively small, we improve the simulation parameters, such as increasing the number of measuring sweeps to 120 000, to ensure accuracy. To achieve data quality comparable to that of cases where $\langle \operatorname{sign}\rangle  \simeq 1.0$, the required runs are estimated to increase by a factor of $\langle \operatorname{sign}\rangle^{-2}$ \cite{santos2003introduction}.
\begin{figure}[h]
	\includegraphics[scale=0.75]{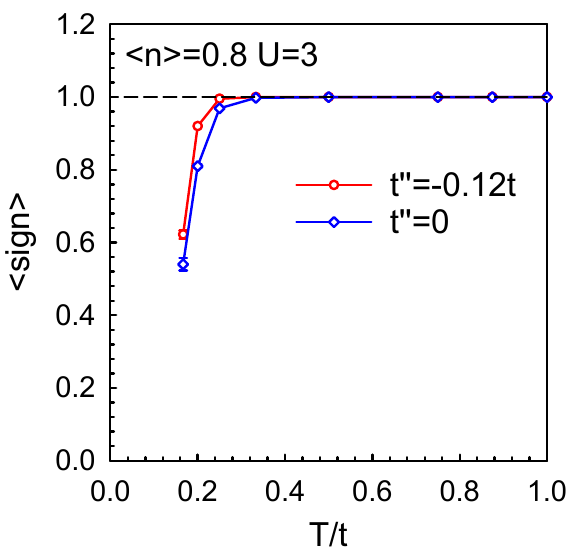}
	\caption{The average fermion sign $\langle \operatorname{sign} \rangle$ as a function of temperature $T$ at $U=3$ and $\langle n \rangle=0.8$ in nickelates and cuprates.}
	\label{FIG9}
\end{figure}

\begin{figure}[h]
	\includegraphics[scale=0.45]{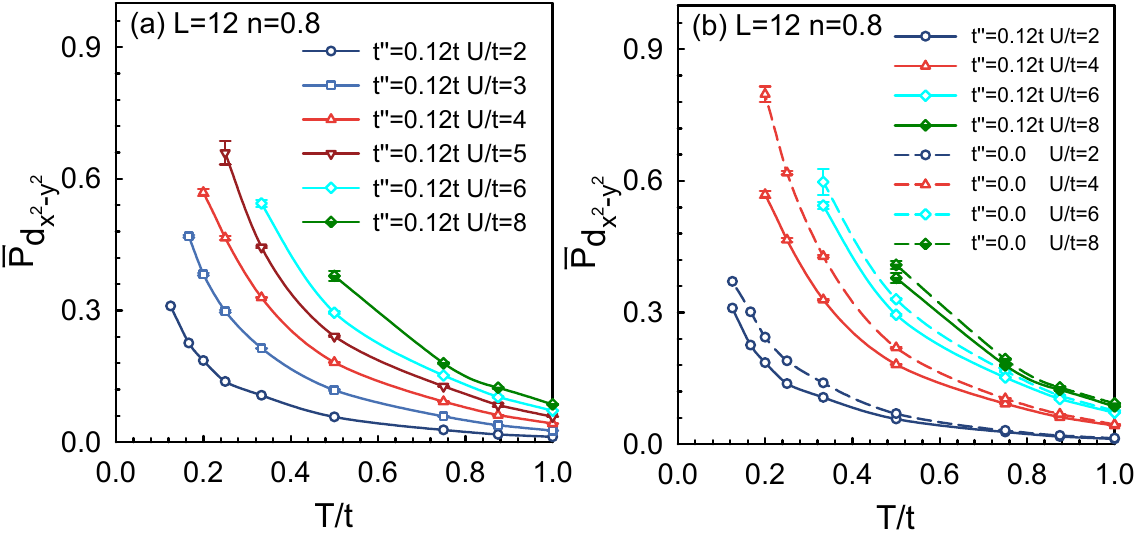}
	\caption{(a) Temperature dependence of the $d_{x^2-y^2}$ effective pairing interaction $\bar{P}_{d_{x^2-y^2}}$ in doped nickelates at different electron densities $\langle n\rangle=0.8$. (b) Temperature dependence of the $d_{x^2-y^2}$-wave pairing susceptibility $\bar{P}_{d_{x^2-y^2}}$ for different $t''$ values at lattice size $L=12$, electron density $\langle n\rangle=0.8$, and interaction strengths $U/t=2,4,6,8$.}
	\label{FIG10}
\end{figure}

Recent theoretical studies using constrained random-phase approximation (cRPA) have revealed that the on-site Coulomb interaction \( U \) in infinite-layer nickelates at ambient pressure is approximately 3.4~eV \cite{di2024unconventional}. It successfully reproduced the experimentally observed increase in critical temperature under pressure up to 12~GPa \cite{wang2022pressure} and have revealed that \( T_c \) could further increase at higher pressures. Theoretical studies have shown that the parameter \( U \) remains largely constant under pressure, due to the limited changes in the orbital shapes and Wannier function spreads. Conversely, the hopping parameter \( t \) increases significantly as Ni atoms move closer together under pressure, increasing orbital overlap at neighboring sites. From ambient pressure to 150~GPa, \( t \) nearly doubles, reducing the \( U/t \) ratio to approximately 5. However, recent resonant inelastic X-ray scattering (RIXS) experiments and theoretical analysis suggest \( U \approx 4.4 \)~eV for nickelates  \cite{PhysRevB.110.224431}, leading to \( U/t \) reaching 7.1 even under high pressure. To explore higher \( U/t \) values, additional calculations for \( U = 5,\, 6,\, \) and \( 8 \) at electron density \( n = 0.8 \) were performed for nickelates and cuprates. Fig \ref{FIG10} shows the temperature dependence of \( \bar{P}_{d_{x^2 - y^2}} \) at these larger \( U/t \) values. The results indicate that, with increasing \( U/t \), \( \bar{P}_{d_{x^2 - y^2}} \) in doped nickelates and cuprates is significantly enhanced. This enhancement is more pronounced at lower temperatures. These findings suggest that the trend observed at smaller \( U/t \) values persists at larger \( U/t \) values: electronic correlation effects play a crucial role in promoting \( d_{x^2 - y^2} \) superconducting pairing. The increase of \( \bar{P}_{d_{x^2 - y^2}} \) with increasing \( U/t \) underscores the importance of strong electron-electron interactions in the high-temperature superconductivity of cuprates and nickelates. Furthermore, the third-nearest hopping \( t'' \) still has a suppressive effect on the \( d_{x^2 - y^2} \) pairing channel, indicating that our results are robust.

\begin{figure}[h]
	\includegraphics[scale=0.6]{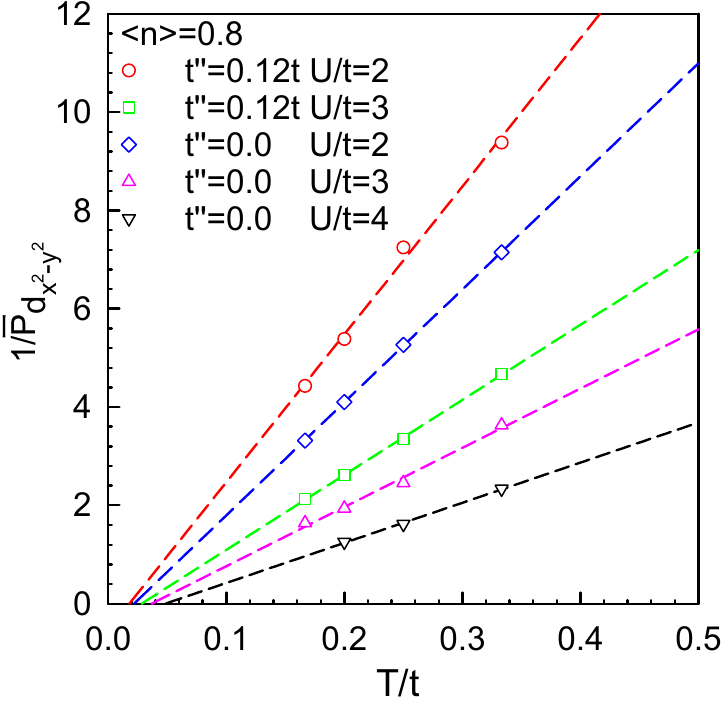}
	\caption{The inverse effective pairing susceptibility \(1/\bar{P}_{d_{x^2-y^2}}\) as a function of temperature for doped nickelates and cuprates at different interaction strengths \(U\).}
	\label{FIG11}
\end{figure}

To further illustrate these trends to lower temperatures, we studied the variation of the inverse effective pairing susceptibility $1/\bar{P}_{d_{x^2-y^2}}$ with temperature for both materials and found a linear correlation between $1/\bar{P}_{d_{x^2-y^2}}$ and $T$, similar to Curie-Weiss behavior $1/\chi  = \left( {T - {T}_{c}}\right) /A$. Therefore, by performing a linear fit and extrapolating to zero temperature, the intersection point of the fitted line with the $T$-axis corresponds to the superconducting transition temperature. Fig. \ref{FIG11} shows that with increasing $U$, the absolute value of the intercept increases, indicating that within the parameter range we studied, superconductivity is indeed driven by strong electron-electron correlations. Additionally, we obtained the difference in the fitted superconducting transition temperatures between nickelates and cuprates when $U=3$. The $T_c$ corresponding to nickelates is $0.028t$, and for cuprates, it is $0.036t$. Their difference is $0.008t$, which is already a considerable gap.

\begin{figure}[h]
	\centering 
	\includegraphics[scale=0.7]{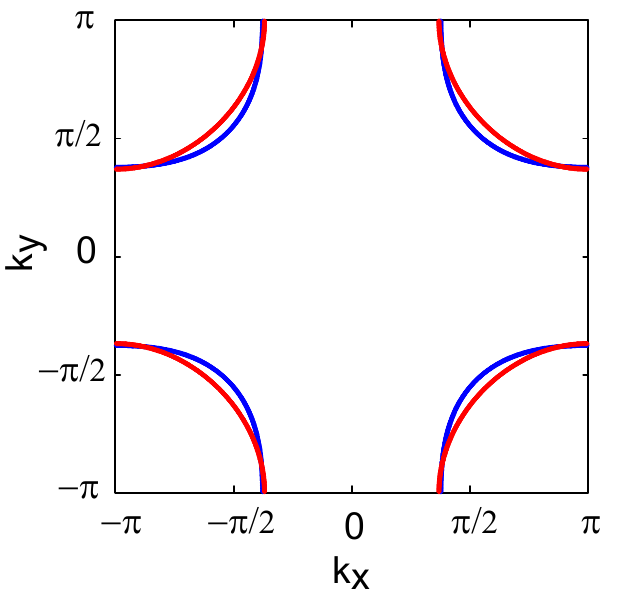}
	
	\caption{Fermi surface for the single-band Hubbard model at 20\% hole doping ($n=0.8$). The blue line represents the Fermi surface with $t^{\prime\prime}=-0.12t$, whereas the red line shows the case of $t^{\prime\prime}=0$. Other parameters are fixed at $t^{\prime}=0.25t$.}
	\label{FIG12}
\end{figure}

The suppression effect mentioned in our main text likely arises because reducing $t''$ correspondingly diminishes the warping of the Fermi surface. Specifically, as a third-nearest hopping term, $t''$ plays a crucial role in modulating Fermi surface warping. In single-band Hubbard models with realistic on-site interactions $U$, a relatively large magnitude of $t''$ tends to suppress superconductivity. Studies based on the fluctuation exchange (FLEX) approximation \cite{bickers1989conserving,dahm1995quasiparticle} have demonstrated that a more rounded Fermi surface, resulting from enhanced third-nearest hopping, leads to a reduction in the superconducting transition temperature $T_c$ \cite{sakakibara2010two}. Conversely, decreasing the magnitude of $t''$ tends to flatten the Fermi surface, reducing its warping and enhancing Fermi surface nesting, which is generally favorable for the emergence of superconductivity. In our investigation of doped nickelates and cuprates, as shown in Fig. \ref{FIG12},  reducing the magnitude of $t''$ within the appropriate parameter range results in a more square-like (or diamond-shaped) Fermi surface, and this geometric modification strengthens the $d_{x^2-y^2}$-wave superconducting pairing. These findings align with recent studies \cite{PhysRevLett.130.166002}, indicating that modulating $t''$ provides an effective means to tune the Fermi surface topology and thereby control the system's superconducting properties.

\bibliography{ref}
	
\end{document}